\scrollmode
\documentstyle[draft]{mn}

\input epsf

\title[Cooling curves]{Cooling curves and initial models for low--mass 
white dwarfs ($< 0.25 ~M_\odot$) with helium core}

\author[Marek J. Sarna et al.]{Marek J. Sarna$\rm ^1$, Ene  
Ergma$\rm ^2$ and Jelena Antipova$\rm ^2$ \\
$\rm  ^1~$ N. Copernicus Astronomical Center, 
       Polish Academy of Sciences,
       ul. Bartycka 18, 00--716 Warsaw, Poland. \\
$\rm ^2~$ Physics Department, Tartu University, \"Ulikooli 18, EE2400 
       Tartu, Estonia. \\}

\date{\small Accepted . Received ; in original 
form 1999 }

\begin{document}

\maketitle

\begin{abstract}

We present a detailed calculation of the evolution of low--mass
($< 0.25~M_\odot $) helium white dwarfs. These white dwarfs (the 
optical companions
to binary millisecond pulsars) are formed via long--term, low--mass
binary evolution. After detachment from the Roche lobe, the hot helium 
cores have a rather thick hydrogen layer with mass between 0.01 to 
0.06$~M_\odot $. Due to mixing between the core and outer envelope, 
the surface 
hydrogen content is 0.5 to 0.35, depending on the initial value of the 
heavy element (Z) and the initial secondary mass. We found that the majority 
of our computed models experience one or two hydrogen shell flashes. 
We found that the mass of the helium dwarf in which the hydrogen shell 
flash occurs depends on the chemical composition. 
The minimum helium white dwarf mass in which a hydrogen flash takes 
place is 0.213$~M_\odot $ (Z=0.003), 0.198$~M_\odot $ (Z=0.01), 
0.192$~M_\odot $ (Z=0.02) or 0.183$~M_\odot $ (Z=0.03).
The duration of the flashes (independent of chemical composition) is between
few $\times 10^6 $ years to few $\times 10^7 $ years. In several flashes 
the white dwarf radius will increase so much that it forces the model to fill
its Roche lobe again.
Our calculations show that cooling history of the helium white dwarf depends
dramatically on the thickness of the hydrogen layer. We show that the
transition from a cooling white dwarf with a temporary stable
hydrogen--burning shell to a cooling white dwarf in which almost all residual
hydrogen is lost in a few thermal flashes (via Roche--lobe overflow) occurs
between 0.183--0.213$~M_\odot $ (depending on the heavy element value). 

\end{abstract}


\section{Introduction}

Kippenhahn, Kohl \& Weigert (1967) were the first who followed the formation
of helium white dwarfs (WD) of low mass in a binary system. The
evolution of a helium WD of 0.26$~M_\odot$ (remnant) was
investigated by Kippenhahn, Thomas \& Weigert (1968) who found that a
hydrogen flash can be initiated near the base of the hydrogen rich
envelope. The energy of the flash is sufficient to cause the envelope
to expand to giant dimensions and hence it may be possible that another
short term Roche lobe filling can occur.

In Webbink (1975), models of a helium white dwarf were constructed
by formally evolving a model from the homogeneous zero--age main sequence 
with the reduction of the mass of the hydrogen--rich envelope. When the 
mass of the envelope is less than some critical value, the model contracts
adopting white dwarf dimensions. Webbink found that thermal flashes do not
occur for WDs less massive than 0.2$~M_\odot$.
Alberts et al. (1996) have confirmed Webbink's finding that 
low--mass white dwarfs do not show thermal flashes and the
cooling age for WDs of mass
$M_{wd}$$\leq$0.20$M_\odot$ can be considerably underestimated if
using the traditional WD cooling curves which were constructed for $
M_{wd}$$>$0.3$M_\odot$ (Iben \& Tutukov 1986, IT 86).
   
Recently, Hansen \& Phinney (1998a -- HP98) and Benvenuto \& 
Althaus (1998 -- BA98) investigated the effect of different mass of
the hydrogen layer 
($\rm 10^{-8} \le M_{env}/M_\odot \le 4 \times 10^{-3} $) on the 
cooling evolution of $\rm 0.15 \le M_{He}/M_\odot \le 0.5 $ helium WDs. In both calculations (BA98 and HP98) the mass of the hydrogen
envelope left on the top of white dwarf has been taken as free parameter.
BA98 found that thick envelopes appreciably modify the radii
and surface gravities of no--H models, especially in the case of low--mass 
helium white dwarfs. 

Driebe et al. (1998 -- DSBH98) present a grid of evolutionary tracks 
for low-mass white dwarfs with helium cores in the mass range from 0.179 to
0.414$~M_\odot$. The tracks are based on a 1$~M_\odot$ model sequence 
extending
from the pre--main sequence stage up to the tip of red giant branch.
Applying large mass loss rates forced the models to move off the giant 
branch and evolve across the Hertzsprung--Russell diagram and down
the cooling branch. They found that hydrogen flashes take place only
for two model sequences, 0.234$~M_\odot$ and 0.259$~M\odot$, and for 
very low--mass WDs 
the hydrogen shell burning remains dominant even down to
effective temperatures well below 10 000 K.
According to our previous calculations (Ergma, Sarna \& Antipova, 1998)
we  find that for a low--mass white dwarf with a helium core, which was formed during low--mass binary evolution (after detachment from the
Roche lobe), the hydrogen layer left on the top of the helium core is much 
thicker ($\rm \sim 1-6 
\times 10^{-2} M_\odot $ with $\rm X_{surf} $ ranging from 0.3 to 0.52) 
than used in cooling calculation by HP98 and BA98. 
Also in DSBH98 (see their Table 1), for the two lowest total remnant 
masses the envelope mass value is smaller that obtained in our calculations. 


\section{The main aim}

Low--mass helium white dwarfs are present in millisecond binary 
pulsars and double degenerate systems. This gives a unique 
opportunity to test the cooling age of the WD in a binary and, 
especially in the case of millisecond binary pulsars, allows for 
age determinations for neutron stars that are independent of 
their rotational history.  

\begin{figure}
\epsfverbosetrue
\begin{center}
\leavevmode
\epsfxsize=7cm
\epsfbox{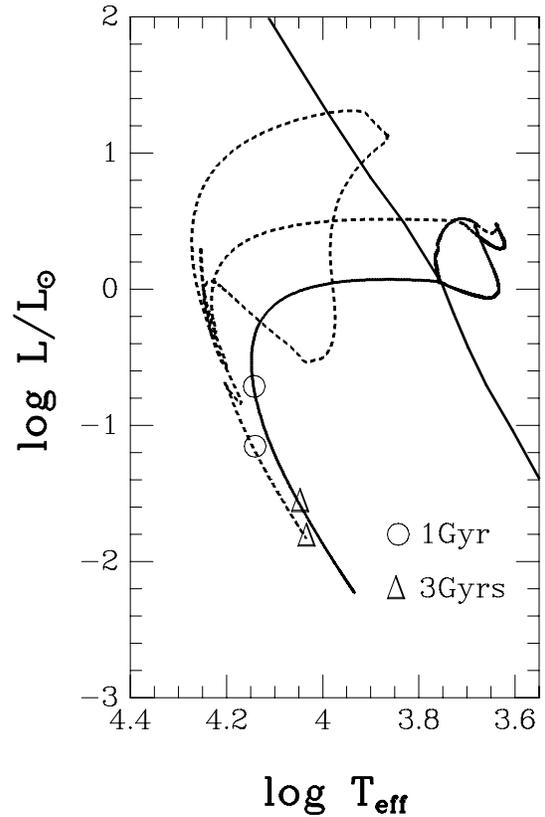}
\end{center}
\caption{Hertzsprung--Russell diagram with evolutionary tracks. Evolutionary 
sequence (model 20) which undergoes long term stable hydrogen burning is 
shown by the solid line. 
Dashed line, the same for model 22 which shows one weak (without RLOF) and 
one strong (with RLOF) hydrogen flash. Circles and triangles mark 
cooling ages of 1 and 3 Gyr, respectively.} 
\end{figure}

\begin{figure}
\epsfverbosetrue\begin{center}
\leavevmode
\epsfxsize=7cm
\epsfbox{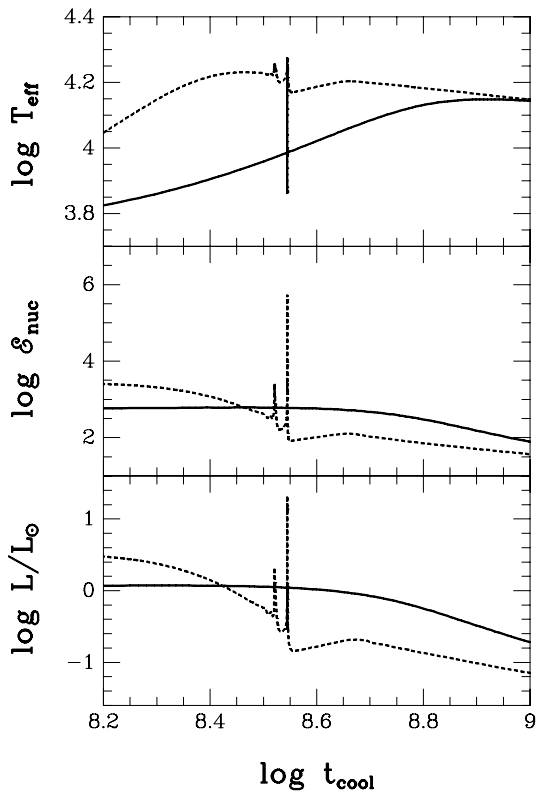}
\end{center}
\caption{The surface effective temperature ({\it upper panel}), the nuclear 
energy production ({\it middle panel}) and the surface luminosity 
({\it lower panel}) plotted as a function of the cooling time $t_{cool}$
which is the time elapsed from $t_{0}$.
Model 20  with stationary hydrogen shell burning -- thick  
line $t_{0}$=7.9$\times 10^{9}$yrs, model 22 with unstable hydrogen shell 
burning -- dashed line $t_{0}$=7.8$\times 10^{9}~$yrs. First
flash is without RLOF and second flash is accompanied by RLOF. For all 
figures with cooling time $t_{cool}$ is time elapsed from $t_{0}$. } 
\end{figure}

\section{The evolutionary code}

The evolutionary sequences we have calculated are comprised of 
three main phases: 

\noindent$\bullet$ detached evolution lasting until the companion 
fills its Roche lobe on the time--scale $t_{d}$; 

\noindent
$\bullet $ semi--detached evolution (non--conservative in our
calculations) on the time--scale $t_{sd}$; $t_{0}$=$t_{d}+t_{sd}$;

\noindent
$\bullet $ a cooling phase of the WD on the time--scale $t_{cool}$ 
(the final phase during which a
system with a ms pulsar + low--mass helium WD is left behind). 
The total evolutionary time is $t_{evol}= t_{0}+t_{cool}$.  

The duration of the detached phase is somewhat uncertain; 
it may be determined either by the nuclear time--scale or by the 
much shorter time--scale of the orbital angular momentum loss owing  
to the magnetized stellar wind. 

In our calculations we assume that the semi--detached evolution of a binary 
system is non--conservative, i.e. the total mass and angular momentum of the 
system are not conserved. We can express the total orbital angular 
momentum (J) of a binary system as

\begin{equation}
{{\dot J} \over J} = \left. {{\dot J} \over J} \right|_{SML} + \left. 
{{\dot J} \over J} \right|_{MSW} + \left. {{\dot J} \over J} \right|_{GR} ,
\end{equation}

where the terms on the right hand side are due to: stellar mass angular
momentum loss from the system, magnetic stellar wind braking, and 
gravitational wave radiation.

\subsection{Stellar mass angular momentum loss}

The formalism which we have adopted is described in 
Muslimov \& Sarna (1993). We introduce the parameter $f_1$ characterizing 
the loss of mass from the binary system and defined by the relations,

\begin{equation}
\dot M = \dot M_2 f_1 ~~~~and~~~~ \dot M_1 
= - \dot M_2 (1 - f_1) , 
\end{equation}

where $\dot M$ is the mass--loss rate from the system, $\dot M_2$ is 
the rate of mass loss from the donor (secondary) star and $\dot M_1$ is 
the accretion rate onto the neutron star (primary). The matter leaving the 
system will carry off its intrinsic angular momentum in agreement with 
formula

\begin{equation}
\left. {{\dot J} \over {J}} \right|_{SML} = f_1f_2 {{M_1 \dot M_2} \over {M_2 M}} 
~~~~~yr^{-1}  , 
\end{equation}

where $M_1$ and $M_2$ are the masses of the neutron star and donor star, 
respectively and M=$M_1$+$M_2$. Here we have introduced the additional 
parameter 
$f_2$, which describes the efficiency of the orbital angular momentum 
loss from the system due to a stellar wind (Tout \& Hall 1991). In our 
calculations we have $f_2$=1 and $f_1$=1; we calculate the fully 
non--conservative case, although additional calculations with $f_1$ 
= 0.9 and 0.5 (with $f_2$=1) give similar results. A similar result to ours 
was found by Tauris (1996), who showed that the change in orbital  
separation due to mass transfer in LMXB (low--mass X-ray binaries) as a 
function of the fraction of exchanged matter
$f_1$ which is lost from system is small (for 0.5$\leq f_1\leq 1$). 
To understand whether the system evolution is conservative 
or non--conservative is not easy in the case of a rapidly rotating 
neutron star; no easy solution can be found. We propose as one 
possibility a factor which may help us to distinguish between the two 
cases -- the surface magnetic field of the neutron star and its evolution 
during the accretion.  

\subsection{Magnetic stellar wind braking}

We also assume that the donor star, possessing a convective envelope, 
experiences magnetic braking (Mestel 1968; Mestel \& Spruit 1987; Muslimov \& 
Sarna 1995), and, as a consequence, the system loses its orbital 
angular momentum. For a magnetic stellar wind we used the formula for 
the orbital angular momentum loss

\begin{equation}
\left. {{\dot J} \over {J}} \right|_{MSW}= -3\times 10^{-7}{{M^2R_2^2} \over 
{M_1M_2 a^5}}~~~~~yr^{-1}  ,
\end{equation}

where $a$ and $R_2$ are the separation of the components and the radius of 
the donor star in solar units.

\subsection{Gravitational wave radiation}

For systems with very short orbital periods, during the 
final stages of their evolution we also take into account the loss of 
orbital angular momentum due to emission of gravitational radiation 
(Landau \& Lifshitz 1971):

\begin{equation}
\left. {{\dot J} \over J} \right|_{GR} = 8.5 \times 10^{-10} {{M_1 M_2 M} 
\over {a^4}}~~~~~yr^{-1} 
\end{equation}

The mass and accompanying orbital angular momentum loss from these system are
poorly understood problems in the evolution of binary stars. As is well known,
the variation of the angular momentum depends critically on the assumed 
model (Ergma et al. 1998). In the case of binary systems with ms pulsar
typically two different models concerning the mass ejection and
angular momentum loss can be adopted. The first is that the amount of
angular momentum lost per 1 gram of ejected matter is equal to the average
orbital angular momentum of 1 gram of the binary. The second is that the 
matter that flows from
the companion star onto the neutron star (after accretion) is ejected 
isotropically with the specific angular momentum of the neutron star. 
In this paper, for our non--conservative approach we have adopted the 
first model. 
This affects significantly our results on the 
semi--detached evolution (see fig. 2 in Ergma et al. 1998), but 
very little changes the cooling time--scale of the helium white dwarf.

\subsection{Illumination of the donor star}

In all cases we have included the effect of illumination of the donor 
star by the millisecond pulsar. In our calculations we assume that 
illumination of the component by the hard (X--ray and $\gamma$--ray) 
radiation from the millisecond pulsar leads to additional heating of its 
photosphere (Muslimov \& Sarna 1993). The effective temperature $T_{eff} $ of the companion during the illumination stage is 
determined from the relation 

\begin{equation}
L_{in} + P_{ill} = 4 \pi \sigma R_2^2 T^4_{eff} , 
\end{equation}

where $L_{in} $ is the intrinsic luminosity corresponding to the 
radiation flux coming from the stellar interior and $\sigma $ is the 
Stefan--Boltzmann constant.

$P_{ill} $ 
is the millisecond pulsar radiation that heats the photosphere, which is 
determined by 

\begin{equation}
 P_{ill} = f_3 \left({{R_2} \over {2a}}\right)^2 L_{rot}
\end{equation}

and $L_{rot}$ is ``rotational luminosity'' of the neutron star due to 
magneto--dipole radiation (plus a wind of relativistic particles)

\begin{equation}L_{rot}= {{2} \over {3 c^3}} B^2 R_{ns}^6 \left({{2 \pi} 
\over {P_p}}\right)^4 ~~~~,
\end{equation}

where $R_{ns}$ is the neutron star  radius, B is the value of the magnetic 
field strength at the neutron star and $ P_p $ is the pulsar period.
$f_3$ is the factor characterizing the efficiency of transformation of 
irradiation flux into thermal energy (in our case we take $f_3 = 2\times 
10^{-3}$). Note that in our calculations 
the effect of irradiation is formally treated by means of modification 
of the outer boundary condition, according to relation (6).

In this paper we do not follow the magnetic field and pulsar period 
($P_p $) evolution, as we did in our earlier papers (Muslimov \& Sarna 1993,
Ergma \& Sarna 1996). We were mainly interested in finding initial models for
low--mass helium white dwarfs and in investigating the initial cooling phase 
of these low--mass helium white dwarfs. From earlier calculations we know that
if the magnetic field strength is greater than about $ 10^9 ~$G, the 
neutron star spins--up to tenths and hundreds of milliseconds, rather than 
several milliseconds. This
leads to a situation where the pressure of the magneto--dipole radiation is
insufficient to eject matter from the system. Also from our 
previous calculations (see for example Ergma \& Sarna 1996) we find that
after accretion of a maximum of about 0.2$~M_\odot $, the neutron star has
spun--up to millisecond periods if B$< 10^9 ~$G. Therefore in this 
paper we accept that after accretion of 0.2$~M_\odot $ the neutron star
spins--up to about 2 ms. After spin--up the pulsar irradiation is strong 
enough to prevent accretion, and at this moment we include non--conservative 
mass loss from the system as described above. 

During the initial high mass accretion phase ($ \dot M_2 \sim 10^{-8} - 
10^{-9} 
~M_\odot ~yr^{-1}$, $t_{acc} \sim 10^7 - 10^8 ~$yrs) the system may be
observed as a bright low--mass X--ray binary (LMXB). It is necessary to point
out that majority of LMXBs for which orbital period determinations are 
available (21 systems out of 24 according to van Paradijs catalogue 1995), 
have orbital period of less than one day. These systems therefore  
cannot be the progenitors of the majority of low--mass helium white dwarf + 
millisecond pulsar binary systems. 
A lack of LMXB systems with orbital period between 1 -- 3 days 
does not allow us to make a direct comparison between the observational 
data and the results of our calculations.

\subsection{The code}

The models of the stars filling their Roche lobes were computed using a 
standard stellar evolution code based on the Henyey--type code of Paczy\'nski 
(1970), which has been adapted to low--mass stars. 
The Henyey method involves iteratively improving a
trail solution for the whole star. During each iteration, corrections to all
variables at all mesh points in the star are evaluated using the
Newton--Raphson method for linearised algebraic equations (see for example
Hansen \& Kawaler 1994). The Henyey method extended to calculate stellar
evolution with mass loss, as adopted here, is well explain by Zi\'o\l kowski
(1970). We note here that our code makes use of the stationary envelope 
technique, which was developed early on in the life of our code in order to
save disc space (Paczy\'nski 1969). This method makes the assumption that the 
surface 0.5 -- 5\% (by mass) of the star is not significantly affected by
nuclear processes, such that it can be treated to a good approximation as 
homogeneous region (in composition) throughout the whole evolutionary
calculation. During the cooling phase we assume that the static envelope 
is the
surface 0.5\% of the star. This assumption is valid during the flashes 
because the time--scale is longer than thermal time--scale of the
envelope. We tested the possibility that 
the algorithm for redistributing meshpoints introduces numerical diffusion 
into the composition profile. 
We find that if such numerical diffusion is real, it has only a marginal 
influence on the hydrogen profile.
We would also like to note that in the heat equation we neglect the 
derivative with respect to molecular weight, since its effect is small. 
Convection is treated with 
the mixing--length algorithm proposed by Paczy\'nski (1969). We solve the 
problem of radiative transport by employing the opacity tables of Iglesias \&
Rogers (1996). Where the Iglesias \& Rogers (1996) tables 
are incomplete, we have filled the gaps using the opacity tables of Huebner et
al. (1977). For temperatures less than 6000 K we use the opacities given by 
Alexander \& Ferguston (1994) and Alexander (private communication). 
The contribution from conduction 
present in the opacity tables of Huebner et al. (1977) has been included by
us in 
the other tables, since they don't include it (Haensel, private 
communication). 
The equation of state (EOS) includes radiation and gas pressure, 
which is composed of the ion and electron pressure. Contribution to the 
EOS owing to the non--ideal effects of Coulomb interaction and pressure 
ionization which influence the EOS, as discussed by Pols et al. (1995), 
have not been included in our program, and for this reason we stopped our 
cooling calculations before these effects become important.
During the initial phase of cooling, the physical conditions in the hot 
white dwarfs are such that these effects are usually small.

\section{Evolutionary calculations}

We perform our evolutionary calculations 
for binary systems initially consisting of a 1.4 $\rm M_{\odot }$ neutron 
star (NS) and a slightly evolved companion (subgiant) of two masses, 
1 and 1.5, and four chemical compositions (Z: 0.003, 0.01, 0.02, 
0.03). We have produced (Table 1)
a number of evolutionary tracks corresponding to the different possible 
values of the initial orbital period (ranging from 0.7 to 3.0 days) at the 
beginning of mass transfer phase.

\section{The Results}

In Table 1 we list the characteristic of the cooling 
phase of the WD, $\rm t_{cool}$, and the maximum possible evolution 
time of a system, $\rm t_{evol}$, which is a sum of times of detached
(determined by nuclear evolution), semi--detached, and cooling phases. 
The cooling is the last phase of evolution of the WD, and in our 
calculations starts at the end of RLOF. 
The cooling
time, $\rm t_{cool}$, is limited to an initial cooling stage during which the 
WD cools until its central
temperature has decreased by 50 $\% $ of its maximum value. From Table 1 it 
is clearly seen that to produce short orbital period 
systems in a time--scale shorter than Hubble time it is necessary either to 
have low Z or a more massive secondary.  

In our calculations the donor star fills its Roche lobe while it is evolving 
through the Hertzsprung gap, and therefore it transfers mass on its companion
in a thermal time--scale. 

Figure 1 show the evolutionary cooling sequences for models 20 and 22 (more 
details in Table 1). Model 20 presents the case with stable hydrogen 
burning.  Model 22  shows the case when the thermal instability of the 
hydrogen--burning shell occurs. The first flash is not strong enough to allow 
the star to overflow its Roche lobe, but during the second flash  the radius of 
the secondary increases to fill its Roche lobe and short--time Roche lobe 
overflow (RLOF) occurs. 

In Table 2 we present the mass--radius relationship for WDs  from our 
calculations, DSBH98, the Wood models, and   
the Hamada \& Salpeter (1961) zero--temperature helium WD models 
calculated for a surface temperature of 8500 K (as in van Kerkwijk, 
Bergeron \& Kulkarini 1996 for PSR1012+5307). Comparison of the numbers demonstrate that for WD 
masses of $<$ 0.25$~M_\odot $, the results of our calculations 
differ significantly from a simple extrapolation obtained from the
cooling curves (Wood 1990) performed for carbon WDs with the thick 
hydrogen envelopes. In addition comparing the cooling 
time--scales of HP98 and BA98 with those of Webbink and our models, shows 
differences of an order of magnitude (Table 3) for WD masses of $<$ 
0.25$~M_\odot $. 

\begin{figure}
\epsfverbosetrue
\begin{center}
\leavevmode
\epsfxsize=7cm
\epsfbox{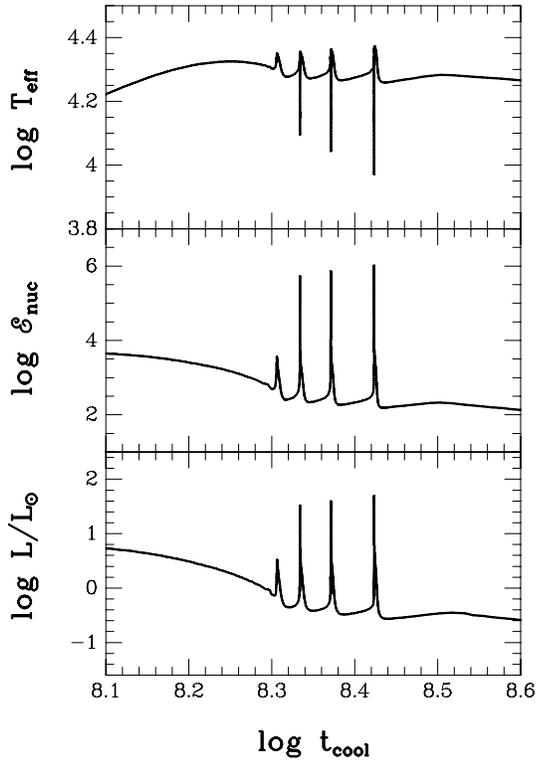}
\end{center}
\caption{Hydrogen flashes on a helium WD of mass 0.213$~M_\odot$
(model 7) which show four flashes without RLOF. The curves present the 
effective temperature ({\it upper panel}), nuclear energy production in the hydrogen burning shell ({\it middle panel}) and the luminosity 
({\it lower panel}) as a function of cooling time, $t_0$=5.2$\times 10^9$ 
yrs.} 
\end{figure}

\begin{figure}
\epsfverbosetrue
\begin{center}
\leavevmode
\epsfxsize=7cm
\epsfbox{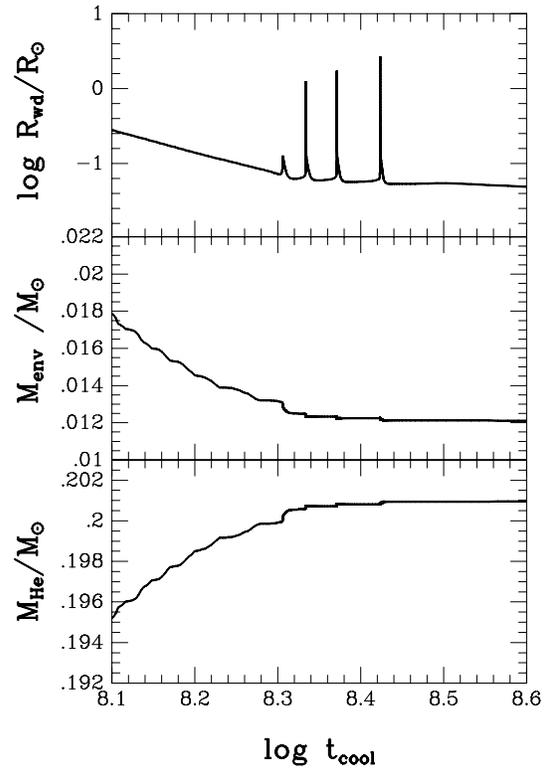}
\end{center}
\caption{Hydrogen flashes on a helium WD of mass 0.213$~M_\odot$ (model 7). 
The curves present the white dwarf radius ({\it upper panel}),  
the envelope mass ({\it middle panel}) and the mass of the helium core 
({\it lower panel}) as a function of the cooling time, $t_0$=5.2$\times 
10^9$yrs.} 
\end{figure}

\begin{figure}
\epsfverbosetrue\begin{center}
\leavevmode
\epsfxsize=7cm
\epsfbox{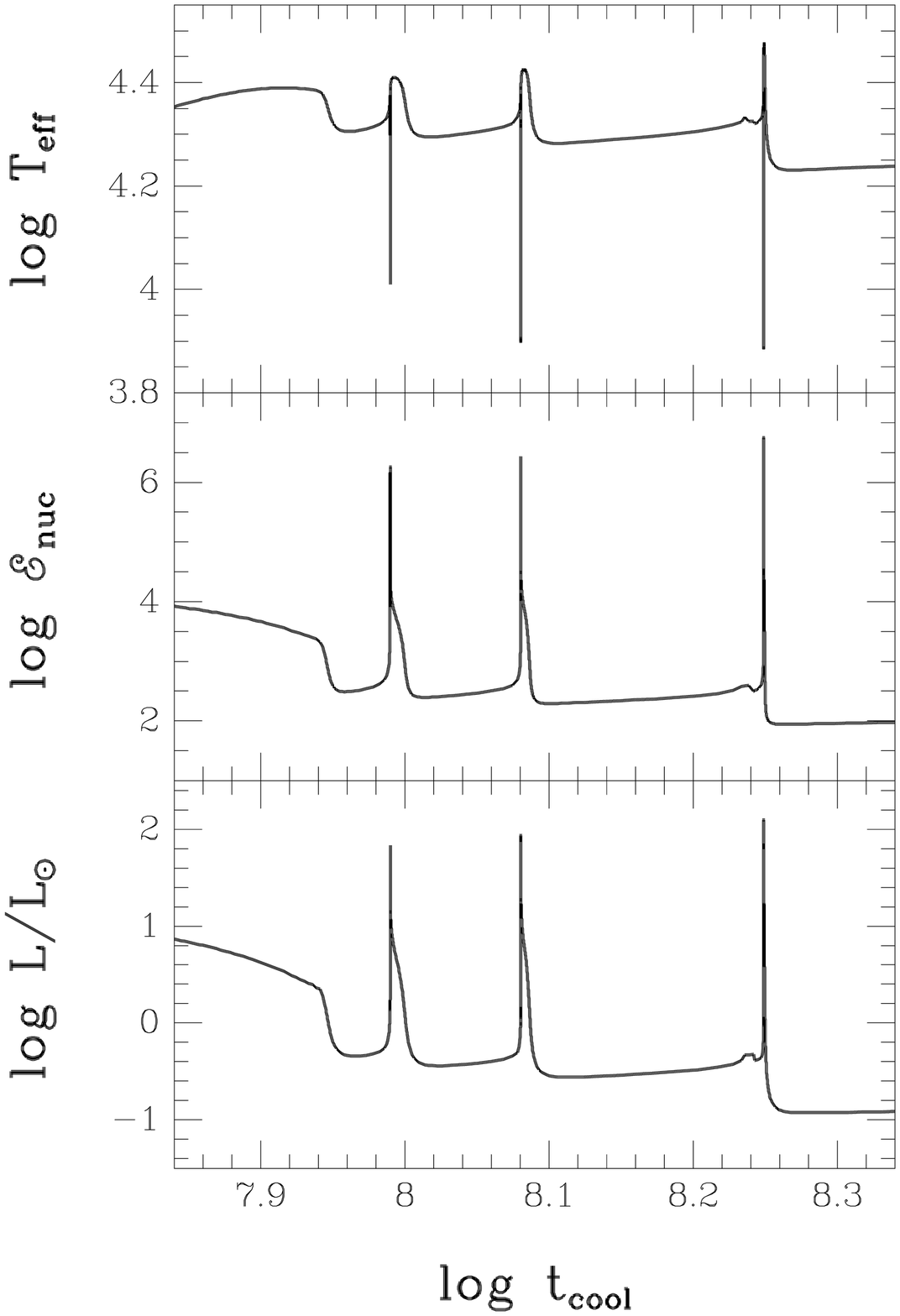}
\end{center}
\caption{Same as for Fig.3 but for model 17. During first flash the secondary 
does not fill its Roche lobe but during the second and third flashes RLOF 
occurs 
and the total mass of white dwarf decreases ($t_0$= 1.4$\times 10^9$ yrs).} 
\end{figure}

\begin{figure}
\epsfverbosetrue
\begin{center}
\leavevmode
\epsfxsize=7cm
\epsfbox{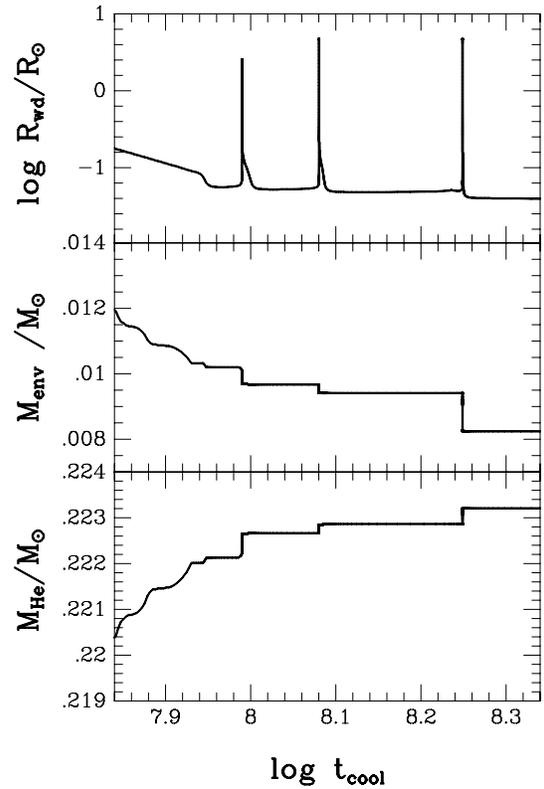}
\end{center}
\caption{Same as for Fig.4 but for model 17. During the first flash 
the secondary does not fill its Roche lobe but during the second and third 
flashes RLOF occurs 
and the total mass of white dwarf decreases ($t_0$= 1.4$\times 10^9$ yrs).} 
\end{figure}

\section{Hydrogen flash burning}

The problem of unstable hydrogen shell burning in low--mass helium WDs 
was first discussed in the literature more than 30 years ago (Kippenhahn, 
Thomas \& Weigert 1968). Recently, Alberts et al. (1996) have claimed that 
they do not 
see any thermal flashes that result from thermally  unstable 
shell--burning, as reported in papers IT86 and Kippenhahn, Thomas \& 
Weigert (1968). Webbink (1975) found that in none of 
his model sequences, such a severe thermal runaway as described 
by Kippenhahn et al. (1968) was found, although mild flashes for 
M$>$0.2 $M_\odot$ did take place. Alberts et al. found that even reducing 
the time step to 50--100 years would not lead to thermally unstable 
shell--burning for $M_{wd}$$<$0.25 $~M_\odot$. In DSBH98, 
thermal instabilities of the hydrogen--burning shell occurs in their 
two models, 0.234$~M_\odot$ and 0.259$~M_\odot$. They concluded that 
hydrogen flashes take place only in the mass interval 0.21$\leq$ $M/M_\odot \leq 0.3$.

According to our computations, low--mass helium WDs with masses more
than 0.183$~M_\odot$ 
(Z=0.03), 0.192$~M_\odot$ (Z=0.02), 0.198$~M_\odot$ (Z=0.01) and 
0.213$~M_\odot$, (Z=0.003) may experience up to several hydrogen 
flashes before they enter the cooling stage. In Table 4  we present several 
characteristics for the computed flashes. We discussion two kinds 
of flashes: in the first case (in Table 4 shown as ``1''), during the
flash the secondary does not fill its 
Roche lobe i.e. the mass of the white dwarf does not change, and in the 
second case  
(``2''), during the unstable hydrogen burning phase the secondary fills its 
Roche lobe and the system again enters into a very short duration 
accretion phase (see Table 4). 
We introduce four time--scales to describe the flash behaviour: (i) the 
flash rise time--scale $\Delta t_1$, which is the time for the luminosity to 
increase from minimum to maximum value (typically this value is between 
few $ \times 10^6$ to few $ \times 10^7$ yrs -- third column in the Table 4);
(ii) the flash decay time--scale $\Delta t_2$, which is the time for the 
luminosity to decrease to the initial value (typically from few hundred thousand to  few tenth million years); 
(iii) $\Delta$ T is the recurrence time between two successive flashes (iv) 
$ \Delta t_{acc}$ is the duration of the accretion phase when the secondary 
fills its Roche lobe during  hydrogen shell flash.

For all sequences with several unstable hydrogen shell burning stages
(usually for case ``1''), the first flash is the weakest. 
In the majority of cases when the flash forces 
the star to fill its Roche lobe, only one flash takes place.  For four cases 
we  found two successive flashes with Roche lobe overflow (models 17, 23, 
24, 31), and for another two cases  (models 47, 53) to the first flash is not 
powerful enough to force the secondary fill its Roche lobe, but during 
the second flash it is.

How does the hydrogen flash  burning influence the cooling time--scale? 
In Fig.2, the luminosity and nuclear energy production rates versus cooling 
time for models 20 and 22 are shown. Model 20 shows stationary hydrogen 
burning and model 22, hydrogen flash burning.
Although before flash model 22 was more luminous than model 20, 
later the situation is reversed. After the flash, 
the burning mass of the 
hydrogen rich envelope in model 22 has decreased to 0.0116$~M_{\odot}$, 
whereas the mass of the hydrogen envelope in model 20, 
in which stationary hydrogen burning occurs, is almost twice as large 
(0.0241$~M_{\odot}$). If  we look at how the maximum  
nuclear energy rate behaves with cooling time, we can see that after the 
flash in model 22,  the maximum energy production rate is less than in 
model 20 (stationary hydrogen burning).
      
In Fig. 3  we present the behaviour of log$ ~T_{eff}$, log $\epsilon_{nuc}$ 
and 
log L/$L_\odot$, and in Fig.4 log $R_{wd}$, $M_{env}$ and $M_f/M_{\odot} $ as 
a function of 
cooling time for model 7. Before the helium white dwarf enters the 
final cooling phase, four unstable hydrogen flash burnings occur.  
The same parameters for model 17 (with RLOF) are shown in Figs. 5 and 6. 

\begin{figure}
\epsfverbosetrue
\begin{center}
\leavevmode
\epsfxsize=7cm
\epsfbox{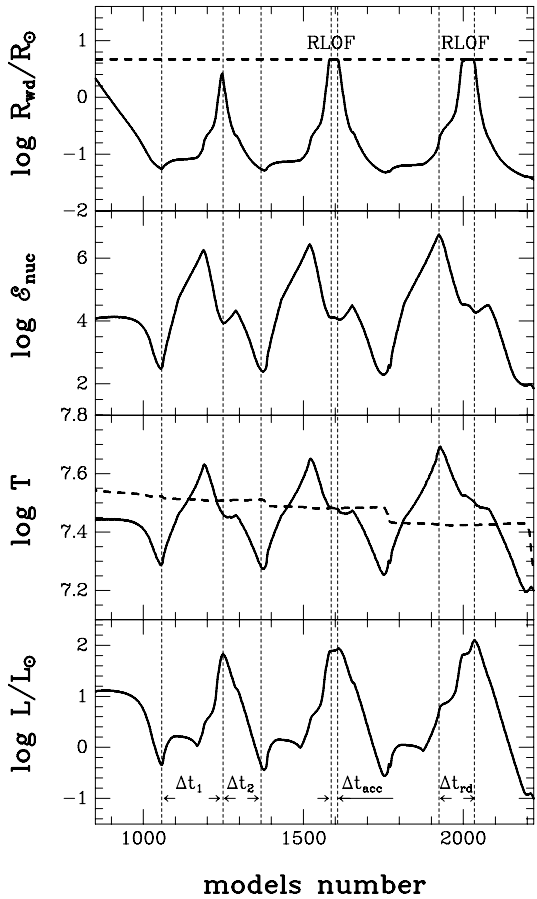}
\end{center}
\caption{Hydrogen flashes on a helium WD of model 17. 
The white dwarf radius (solid line) together with Roche lobe radius (dashed
line) ({\it upper panel}) the nuclear energy production in the hydrogen 
burning shell ({\it upper middle panel}) the maximum shell temperature 
(solid line) and central temperature (dashed line) ({\it lower middle panel})
and the surface luminosity ({\it lower panel}) as a function of model 
number are shown. The vertical lines define different time--scales 
during the flashes.} 
\end{figure}

To investigate in more detail how the flashes develop, we show in Fig. 7 
the evolution of the white dwarf radius (upper panel), nuclear energy 
generation rate (upper middle panel), maximum shell
temperature and central temperature (lower middle panel) and the surface 
luminosity (lower panel) as a function of computed model number. 
In Fig. 7, as vertical dashed lines we marked several time--scales which
characterize the flash behaviour (for numbers see Table 4). $\Delta t_1$ and 
$\Delta t_2 $ describe the rise and decay times; the first 
characterizes the nuclear shell burning time--scale ($ \tau^{shell}_{nuc}$), 
the second the Kelvin--Helmholtz (thermal) envelope time--scale 
modified  
by nuclear shell burning ($\Delta t_2 = \sqrt {\tau^{env}_{K-H}
\tau^{shell}_{nuc}} $). The accretion time ($\Delta t_{acc} $) is described
by the square of the Kelvin--Helmholtz time--scale. The radiative diffusion 
time is defined as the Kelvin--Helmholtz time--scale of the extended envelope 
above the shell ($\Delta t_{rd} = \tau^{env}_{K-H} $). 
The shape of the first flash on Fig. 7 
shows some characteristic changes which are connected with physical 
processes in the stellar interior. At the beginning of the flash 
the luminosity increases due to the more effective hydrogen burning 
in the shell source. After reaching a local maximum, the luminosity then 
decreases while the nuclear energy generation rate is still increasing 
rapidly. This decrease of the surface luminosity is due to a temperature 
inversing forming below the hydrogen shell. The energy generated in 
the hydrogen shell splits into two fluxes; coming outwards and going inwards. 
The helium core is heated effectively by the shell nuclear source -- the 
central temperature increases by 2\%. On Fig. 8 the evolution of the 
luminosity and temperature profiles during the $\Delta t_1 $ and 
$\Delta t_2 $ phases are shown. We clearly see how the inversion profile 
evolves and how the luminosity wave moves into the surface.

\begin{figure}
\epsfverbosetrue
\begin{center}
\leavevmode
\epsfxsize=7cm
\epsfbox{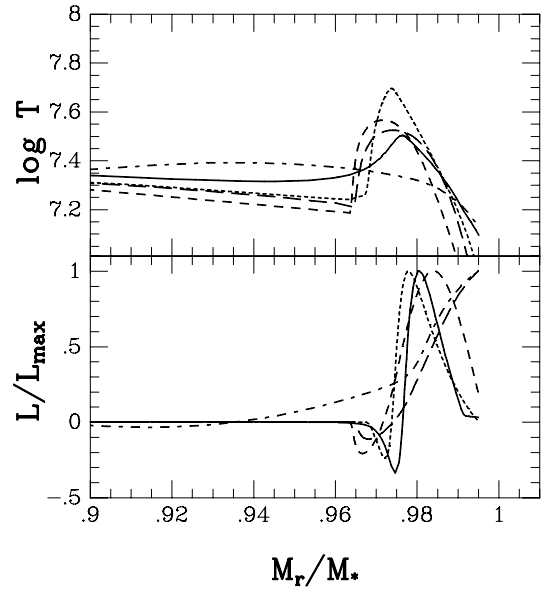}
\end{center}
\caption{The evolution of temperature inversion layers and luminosity 
profile during a hydrogen flash. The evolutionary sequences are as follow: 
solid line -- local luminosity minimum; dashed line -- maximum temperature of
the hydrogen shell; short dashed line -- luminosity front moves outwards; 
long dashed line -- maximum luminosity; dashed doted line -- decline of
luminosity, heated core cooling effectively.} 
\end{figure}
 
The nuclear energy generation rate in the shell has a maximum value far 
away from maximum surface luminosity. This is because the luminosity front 
is moving towards the stellar surface in a time--scale described by radiative 
diffusion ($\Delta t_{rd} $). After reaching a maximum value, the luminosity 
starts to decrease and the 
energy generation rate also declines in the hydrogen shell over a time-scale  
$\Delta t_2 $ (for a contracting envelope) the 
luminosity decreases to the minimum value. During the first flash, 
the stellar radius does not fill the inner Roche lobe. In the second and 
third flashes we have short
episodes of super--Eddington mass transfer (see Table 4). During the RLOF 
phase, the orbital period slightly increases and the subgiant companion 
evolves quickly from spectral type F0 to A0.

As already pointed out, for several cases the secondary 
fills its Roche lobe and the system enters an accretion phase. 
During RLOF, the mass accretion rate is about three orders of magnitude
greater than the Eddington limit (Fig. 9). All the accreted matter 
will be lost from the system ($\Delta M_{acc} \sim 0.0001 - 0.001 ~M_\odot $). 
The accretion phase is very short, usually less than 1000 years (ranging 
from 160 to 2500 yrs -- see Table 4). During the short super--Eddington 
accretion phase the system is a very bright X--ray source, with orbital period
between 2 to 8 days.

We notice that during the flash the evolutionary time step strongly decreases 
and may be as short as several years.

\begin{figure}
\epsfverbosetrue
\begin{center}
\leavevmode
\epsfxsize=7cm
\epsfbox{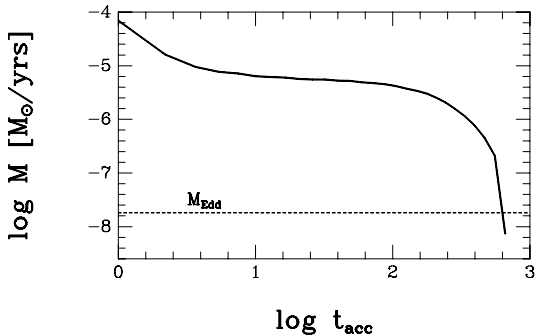}
\end{center}
\caption{Mass accretion rate (model 42) versus time during a hydrogen shell 
flash with RLOF} 
\end{figure}

\section{Role of binarity in the cooling history of the low--mass white dwarf}

DSBH98 modelled single star evolution and produced white dwarfs with various 
masses by applying large mass loss rates at appropriate positions in the 
red--giant 
branch to force the models to move off the giant branch. To show how 
binarity influences the final fate of the white dwarf cooling, 
we have computed  extra 
sequences  (1.0+ 1.4 $M_\odot$, Z=0.02, $P_i$=2.0 days) where we did not 
take into account that the star is in a binary system e.g. during hydrogen 
shell flash we do not allow RLOF. In complete binary model calculation, 
only one shell 
flash occurs accompanied with RLOF, whereas for the single star model 
calculation, four 
hydrogen shell flashes take place.  Due to RLOF, the duration 
of the flash phase is 2.7$\times 10^6 $yrs; if we do not include binarity 
the duration of the flash phase is 1.8$\times 10^8 $yrs. However, 
the cooling time for helium white dwarfs 
less massive than 0.2$~M_\odot $ is not significantly changed. 
This is because the duration of flash phase 
is very short in comparison to the normal cooling phase 
(towards the white dwarf region). However, the effect of binarity 
will be important for the cooling history of more massive helium white
dwarfs. In Fig. 10  both cases of evolution on the Hertzsprung--Russell 
diagram are shown -- on the left panel Roche lobe overflow is not allowed, 
on right panel RLOF takes place.         

\section{Application to individual systems}

Below we discuss the observational data for several systems for which 
results of our calculations may be applied, by taking into account the 
orbital parameters of the system, the pulsar spin--down time, and the white 
dwarf cooling timescale.  

\subsection{\bf PSR J0437--4715}

Timing information for this millisecond binary system: $P_p$=5.757 ms, 
$P_{orb}$=5.741 days, $\tau$ (intrinsic characteristic age of pulsar) = 
4.4 -- 4.91 Gyrs, mass function f(M)= 1.239$10^{-3}$$M_\odot$ (Johnston et 
al. 1993; Bell et al. 1995). Hansen \& Phinney (1998b) have discussed the 
evolutionary stage of this system using their own cooling models described 
in HP98. They found consistent solution for all masses in the range 0.15 -- 
0.375$~M_\odot$ with thick (in the terminology of HP98) hydrogen envelopes 
of 3$\times 10^{-4} ~M_\odot$.

\begin{figure*}
\epsfverbosetrue
\begin{center}
\leavevmode
\epsfxsize=7cm\epsfbox{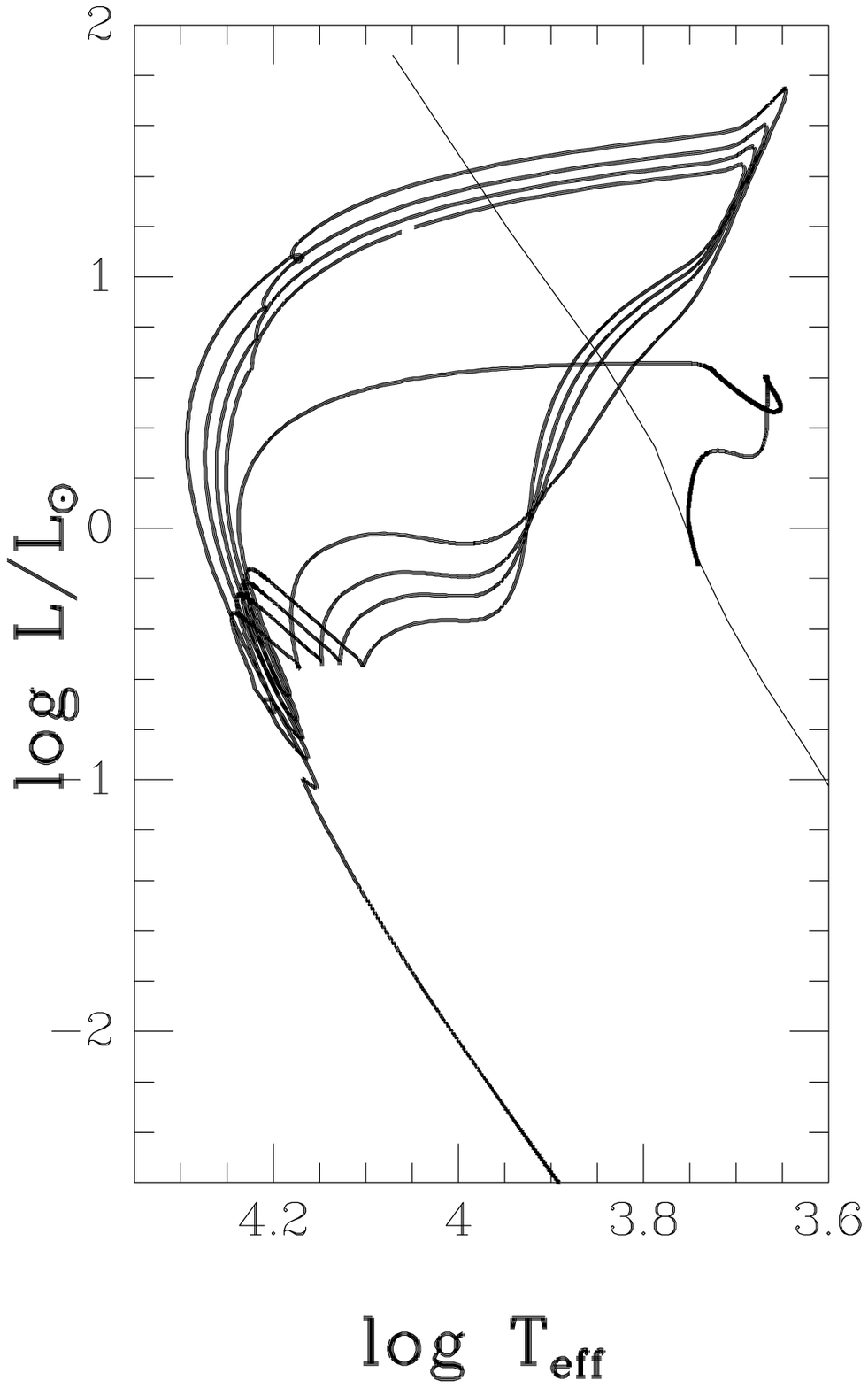}\quad
\epsfxsize=7cm\epsfbox{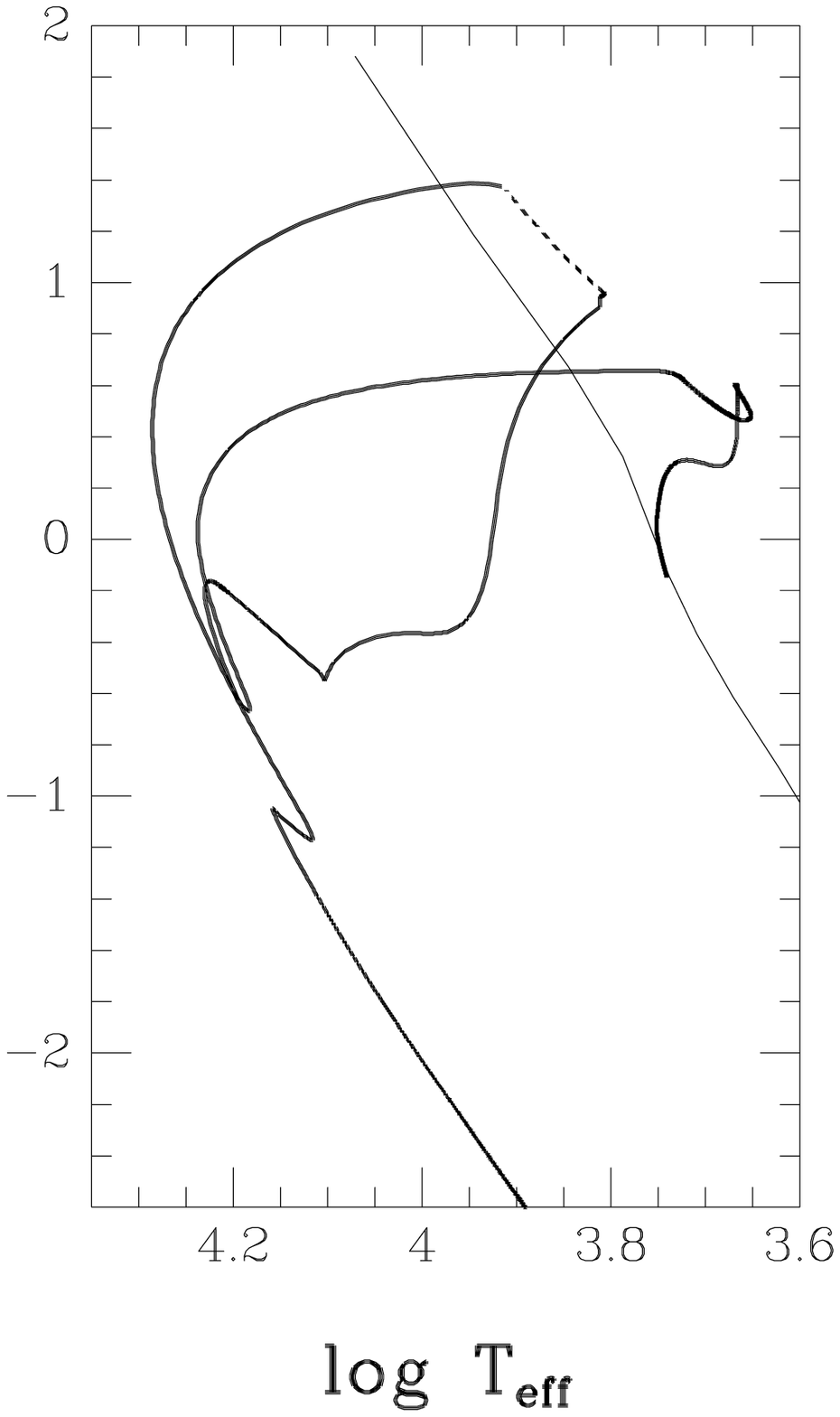}
\end{center}
\caption{Hertzsprung--Russell diagram with evolutionary tracks. 
Evolutionary sequence 1+1.4 $M_\odot$, Z=0.02, $P_i$=2.0 days. Left 
panel RLOF is not allowed, right panel with RLOF.} 
\end{figure*} 

Timing measurements by Sandhu et al. (1997) have detected a rate of change 
in the projected orbital separation $\it a \sin i$, which they interpret 
as a change in $\it i$ and they calculate for an upper limit for 
$\it i$$<$ $43^0$ 
and new lower limit to the mass of the companion of M$\sim$ 0.22$~M_\odot$. 
Our calculations also allow us to produce the orbital parameters and 
secondary mass for the PSR 
J0437--4715 system and fit its cooling age (2.5--5.3 Gyrs, Hansen \& Phinney, 
1998b), and we find that the secondary fills its Roche lobe when the orbital period 
$P_i$ is $\sim$ 2.5 days (Tables 1, 4). From our cooling tracks for 
a binary orbital
period of 5.741 days, the mass of the companion is 0.21$\pm 0.01 ~M_\odot $ 
and its cooling age 1.26--2.25 Gyrs (for a Population I chemical 
composition). These cooling models usually have one strong (with RLOF) 
hydrogen shell flash, after which the helium WD enters the normal cooling 
phase. 

\subsection{\bf{PSR J1012+5307}}

Lorimer et al. (1995) determined a characteristic age of the radio pulsar to 
be 7 Gyr, which could be even larger if the pulsar has a significant 
transverse velocity (Hansen \& Phinney 1998b). Using the IT86 cooling 
sequences, they estimated the companion to be at most 0.3 Gyr old. HP98 
models yield  the following results for this system: the companion mass 
lies in the range 0.13--0.21$~M_\odot$ and the WD age is $<$ 0.6 Gyr, 
the neutron star mass in the range 1.3--2.1$~M_\odot$.

Alberts et al. (1996) were the first to show that the cooling timescale of a 
low--mass 
WD can be substantially larger if there are no thermal flashes which 
lead to RLOF and a reduction of the hydrogen envelope mass. Our and DSBH98 
calculations confirmed their results that for low--mass helium WDs ($<$ 0.2 
$M_\odot$), indeed stationary hydrogen burning plays  important role. To 
produce short (less that one day) orbital period systems with a low--mass 
helium WD and a millisecond pulsar it is necessary that the secondary fills 
its 
Roche lobe between $P_{bif}$ and $P_b$ (Ergma, Sarna \& Antipova, 1998). 
If the initial orbital period $P_{i}$ (at RLOF) is less than $P_{bif}$, 
the binary system evolves towards short orbital periods. $P_b$ is another 
critical orbital period value. If $P_b$ $<$ $P_{i}(RLOF)$ $<$ $P_{bif}$, then 
a short orbital period ($<$ 1 day) millisecond binary pulsar with 
low--mass helium white dwarf may form. 
So the initial conditions of the formation of  
such systems are rather important. We calculated one extra sequence to 
produce a binary system with orbital parameters similar to PSR J1012+5307. 
Initial system: 1 + 1.4 $M_\odot$, $P_{i}$(RLOF) = 1.35 days, Z=0.01. 
Final system : $M_s$=0.168$~M_\odot$, $P_f$=0.605 days, 
$M_{env}$=0.041$~M_\odot$. 
In Fig. 11 in the effective temperature and gravity diagram we 
show the cooling history of this white dwarf after 
detachment of the Roche lobe. The two horizontal  regions are the gravity 
values inferred by van Kerkwijk et al. (1996) (lower) and  Callanan et al. 
(1998) (upper). Our results are consistent with the Callanan et al. (1998) 
estimates. It is necessary to mention that after detachment from its 
Roche lobe, the outer envelope is rather helium--rich. Bergeron et al. (1991)
have shown that a small amount of helium in a hydrogen--dominated envelope 
can mimic the effect of a larger gravity.

\section{Discussion}

The results of our evolutionary calculations differ 
from those of Iben \& Tutukov (1986) and Driebe et al. (1998) because 
of the different formation scenarios for 
low--mass helium WDs. In IT86's calculations a donor star 
fills its Roche lobe while it is on the red giant branch (i.e. has a thick
convective envelope) with a well developed helium core and a thin hydrogen 
burning 
layer. They proposed that the mass transfer time scale is so short that the 
companion will not be able to accrete the transferred matter and will itself 
expand and overflow its Roche lobe. The final output is the formation of a 
common envelope and
the result of this evolution is a close binary with a helium WD
of mass $\rm 0.298~M_{\odot }$ having a rather thin ($\rm 1.4 \times 
10^{-3}~M_{\odot }$) hydrogen--rich (X=0.5) envelope.

DSBH98 did not calculate the mass exchange phases during the red giant 
branch evolution in detail but they also simulated the mass--exchange 
episode by subjecting a red giant branch model to a sufficiently large mass 
loss rate. In both cases (IT86 and DSBH98) mass loss starts when the star 
(with a well developed helium core) is on the red giant branch.

In our calculations the Roche lobe overflow starts when the secondary 
has either almost exhausted hydrogen in the center of the star or 
has a very small helium core with a thick hydrogen burning layer. During 
the semi--detached evolution the mass of the helium core increases from almost 
nothing to final value (for more detail about evolution of such systems, see 
Ergma, Sarna \& Antipova, 1998). 
This is the reason that a much thicker ($\rm \sim [1.5-6] \times 
10^{-2}~M_{\odot }$, with X ranging from 0.30 to 0.52) hydrogen--rich
envelope is left on the donor star at the moment it shrinks 
within the Roche lobe.

The second important point where our results differ from that of DSBH98 is 
that in our calculations we can produce (after the secondary detaches 
from its 
Roche lobe) final millisecond binary pulsar parameters which we  
compare with observational data (orbital period, 
spin period of ms pulsar, mass of the companion). It was shown by 
Joss, Rappaport \& Lewis (1987) and more recently by 
Rappaport et al. (1995) that the evolution of a binary system initially 
comprising of a neutron star and a low--mass giant will end up as a wide  binary 
containing a radio pulsar and a white dwarf in a nearly circular orbit. The 
relation between the white dwarf mass and orbital period (see eq. (6) in 
Rappaport et al. 1995) shows that if the secondary fills its Roche lobe 
while on the red giant branch, then for $M_{wd} \approx 0.19M_\odot$ 
the final orbital period would 
be $\sim$ 5 days, which is far from observed orbital period of the binary 
pulsar PSR J 1012+ 5307 ($P_{orb}$=0.6 days).

Alberts et al. (1996), DSBH98, and the results of our calculations 
demonstrate 
clearly that especially for low--mass helium WDs ($<$ 0.2$~M_\odot$) 
stationary hydrogen burning remains an important, if not the main, 
energy source. HP98 and BA98 did consider nuclear 
burning but found it to be of little importance since their artificially chosen  
hydrogen envelope mass was less than some critical value, disallowing 
significant hydrogen burning. If we compare now the cooling curves of 
HP98, DSH98 with ours then there is one very important difference; 
they did not model the evolution of the helium WD progenitor and all their 
cooling models (see for example  Figs. 11, 12 in HP98) start with a high 
$T_{eff}$. In 
our models, cooling of the helium WD starts after detachment of the secondary 
from its Roche lobe (DSBH98 mimic this situation with mass loss from the 
star). This time, the secondary (proto--white dwarf) has rather low effective 
temperature (see for example Fig.1). During the evolution with L approximately 
constant, the effective temperature increases to a maximum value, after which 
it decreases while still having a active hydrogen shell burning source. 
The evolutionary time needed for the proto--white dwarf to travel from the 
minimum $T_{eff}$ (after detachment from Roche lobe) to maximum $T_{eff}$ 
depends strongly on mass of the WD (for a smaller mass a longer evolutionary 
time--scale).

So for low--mass helium WDs the evolutionary prehistory plays a very 
important role in cooling history of the white dwarf.

\begin{figure}
\epsfverbosetrue
\begin{center}
\leavevmode
\epsfxsize=7cm
\epsfbox{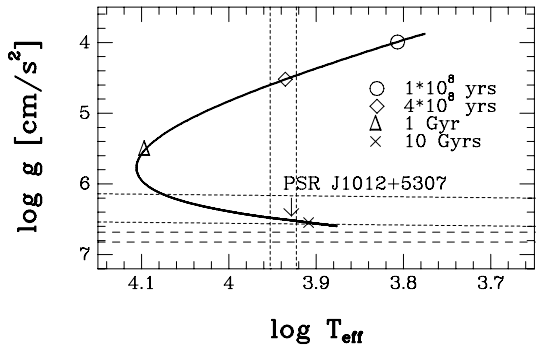}
\end{center}
\caption{log g -- log $T_{eff}$ diagram with $M_{wd}$= 0.168$~M_\odot$. 
The arrow marks the position of the PSR J1012+5307 white dwarf. Two horizontal
regions are the gravity values inferred by Callanan et al. (1998) (upper) 
and van Kerkwijk et al. (1996) (lower). The vertical lines show effective 
temperature constraints of Callanan et al. (1998)}
\end{figure}    

\section{Conclusion}

We have performed comprehensive evolutionary calculations to
produce a close binary system consisting of a NS and a low--mass helium
WD. 

We argue that the presence of a thick hydrogen layer changes 
dramatically the cooling
time--scale of the helium white dwarf ($\rm < 0.25 ~M_\odot $), compared to
the  previous calculations (HP98, BA98) where the mass of the hydrogen 
envelope 
was chosen as free parameter and was usually one order of magnitude
less than that obtained from real binary evolution computations. 

Also, we have demonstrated that using new cooling tracks 
we can consistently explain the evolutionary status of the binary pulsar PSR 
J1012+53. 

Tables with cooling curves are available on {\bf 
http://www.camk.edu.pl/~sarna/}.

\section*{\sc Acknowledgments}

We would like to thank Dr. Katrina M. Exter for help in improving the form
and text of the paper. We would like to thank our referee Dr. Peter Eggleton 
for 
very useful referee opinion. At Warsaw, this work is supported through 
grants 2--P03D--014--07 and 2--P03D--005--16 of the 
Polish National Committee for Scientific Research. Also, J.A. and E.E. 
acknowledges support through Estonian SF grant 2446.

\begin{table*}
\begin{center}
\begin{tabular}{clrrrrrlrrcc}
\multicolumn{12}{c}{Table 1a Cooling track characteristics} \\
\multicolumn{12}{c}{} \\
\hline
\multicolumn{12}{c}{} \\
model & {$ P_{i}$} & \multicolumn{1}{c}{$M_{i}$} & {$\lg t_{cool}$}
& \multicolumn{1}{c}{$\lg t_{evol}$} & {$X_{surf}^{f}$} & 
\multicolumn{1}{c}{$P_{f}$} & \multicolumn{1}{c}{$M_{f}$} & {$M_{2,He}(RLOF)$} 
& {$M_{2,He}(COOL)$} & {$\lg L_{f}$} &  {$\lg T_{eff,f}$} \\
& {[days]} & [{$M_\odot$}] & \multicolumn{1}{c}{[yrs]} 
& \multicolumn{1}{c}{[yrs]} & & {[days]} & [{$M_\odot$}] 
& \multicolumn{1}{c}{[$M_\odot$]} & 
\multicolumn{1}{c}{[$M_\odot$]} & [{$L_\odot$}] & \\
\multicolumn{12}{c}{} \\
\hline
\multicolumn{12}{c}{} \\
\multicolumn{12}{c}{Z=0.003} \\
\multicolumn{12}{c}{} \\

1 & 1.02 & 1.0 & 10.004 & 10.197 & 0.38 & 0.421 & 0.172 & 0.112 & 0.150 
& -2.299 & 3.920 \\
2 & 1.05 & 1.0 & 9.948 & 10.159 & 0.39 & 0.554 & 0.175 & 0.115 & 0.152 
& -2.232 & 3.936 \\
3 & 1.10 & 1.0 & 9.867 & 10.108 & 0.39 & 0.708 & 0.178 & 0.119 & 0.158 
& -2.139 & 3.959 \\
4 & 1.30 & 1.0 & 9.685 & 10.008 & 0.40 & 1.180 & 0.187 & 0.134 & 0.169 
& -1.944 & 4.008 \\
5 & 1.50 & 1.0 & 9.582 & 9.958 & 0.40 & 1.584 & 0.192 & 0.143 & 0.177 
& -1.843 & 4.035 \\
6 & 2.00 & 1.0 & 9.399 & 9.885 & 0.41 & 2.614 & 0.203 & 0.160 & 0.189 
& -1.673 & 4.080 \\ 
7 & 2.50~~H & 1.0 & 9.211 & 9.831 & 0.43 & 4.275 & 0.213 & 0.176 & 0.201 
& -1.499 & 4.125 \\
8 & 3.00~~H & 1.0 & 9.107 & 9.809 & 0.44 & 5.498 & 0.219 & 0.185 & 0.207 
& -1.403 & 4.142 \\
\multicolumn{12}{c}{} \\

9 & 0.70 & 1.5 & 9.479 & 9.645 & 0.43 & 1.591 & 0.191 & 0.137 & 0.175 & -1.659
& 4.066 \\
10 & 0.80 & 1.5 & 9.122 & 9.474 & 0.44 & 2.092 & 0.199 & 0.146 & 0.184 & -1.364
& 4.129 \\
11 & 0.90 & 1.5 & 9.061 & 9.392 & 0.44 & 2.450 & 0.204 & 0.154 & 0.190 
& -1.201 & 4.165 \\
12 & 1.20~~H* & 1.5 & 8.845 & 9.304 & 0.46 & 3.409 & 0.213/ & 0.169 & 0.200 
& -0.986 & 4.210 \\
& & & & & & & 0.212 & & & & \\
13 & 1.50~~H* & 1.5 & 8.788 & 9.286 & 0.48 & 4.280 & 0.217/ & 0.178 & 0.206 
& -0.929 & 4.224 \\
& & & & & & & 0.217 & & & & \\
14 & 1.80~~H* & 1.5 & 8.753 & 9.277 & 0.49 & 5.105 & 0.221/ & 0.185 & 0.210 
& -0.925 & 4.230 \\
& & & & & & & 0.221 & & & & \\
15 & 2.10~~H* & 1.5 & 8.766 & 9.283 & 0.49 & 5.866 & 0.225/ & 0.192 & 0.214 
& -0.957 & 4.229 \\
& & & & & & & 0.224 & & & & \\
16 & 2.50~~H* & 1.5 & 8.742 & 9.278 & 0.48 & 6.831 & 0.229/ & 0.197 & 0.219 
& -0.976 & 4.232 \\
& & & & & & & 0.228 & & & & \\
17 & 3.00~~H* & 1.5 & 8.665 & 9.259 & 0.49 & 7.888 & 0.232/ & 0.203 & 0.223 
& -1.010 & 4.231 \\
& & & & & & & 0.231 & & & & \\
\multicolumn{12}{c}{} \\
\hline
\multicolumn{12}{c}{} \\
\multicolumn{12}{c}{Z=0.01} \\
\multicolumn{12}{c}{} \\
18 & 1.30 & 1.0 & 10.212 & 10.388 & 0.38 & 0.366 & 0.163 & 0.120 & 0.143 
& -2.601 & 3.847 \\
19 & 1.35 & 1.0 & 9.907 & 10.316 & 0.39 & 0.605 & 0.168 & 0.127 & 0.150 
& -2.477 & 3.877 \\
20 & 1.45 & 1.0 & 9.886 & 10.193 & 0.40 & 1.092 & 0.177 & 0.139 & 0.161 
& -2.231 & 3.934 \\
21 & 1.65 & 1.0 & 9.661 & 10.094 & 0.42 & 1.945 & 0.188 & 0.154 & 0.175 
& -1.995 & 3.993 \\
22 & 2.00~~H* & 1.0 & 9.490 & 10.037 & 0.43 & 2.936 & 0.197/ & 0.166 & 0.185 
& -1.829 & 4.035 \\
& & & & & & & 0.196 & & & & \\
23 & 2.50~~H* & 1.0 & 9.165 & 9.967 & 0.45 & 4.272 & 0.205/ & 0.173 & 0.194 
& -1.606 & 4.085 \\
& & & & & & & 0.203 & & & & \\
24 & 3.00~~H* & 1.0 & 9.152 & 9.965 & 0.45 & 5.546 & 0.211/ & 0.184 & 0.201 
& -1.546 & 4.104 \\
& & & & & & & 0.209 & & & & \\
\multicolumn{12}{c}{} \\

25 & 0.90 & 1.5 & 9.902 & 10.007 & 0.44 & 1.075 & 0.174 & 0.132 & 0.156 
& -2.168 & 3.937 \\
26 & 1.05 & 1.5 & 9.650 & 9.810 & 0.47 & 1.855 & 0.186 & 0.148 & 0.172 
& -1.872 & 4.008 \\
27 & 1.10 & 1.5 & 9.596 & 9.772 & 0.46 & 2.032 & 0.188 & 0.152 & 0.175 
& -1.832 & 4.020 \\
28 & 1.20 & 1.5 & 9.504 & 9.710 & 0.47 & 2.378 & 0.192 & 0.157 & 0.180 
& -1.741 & 4.042 \\
29 & 1.50~~H* & 1.5 & 9.368 & 9.629 & 0.47 & 3.152 & 0.200/ & 0.169 & 0.188 
& -1.645 & 4.069 \\
& & & & & & & 0.199 & & & & \\
30 & 2.00~~H* & 1.5 & 9.273 & 9.578 & 0.48 & 4.153 & 0.206/ & 0.178 & 0.195 
& -1.572 & 4.091 \\
& & & & & & & 0.205 & & & & \\
31 & 2.50~~H* & 1.5 & 9.111 & 9.505 & 0.49 & 5.091 & 0.211/ & 0.185 & 0.201 
& -1.475 & 4.114 \\
& & & & & & & 0.209 & & & & \\
32 & 3.00~~H* & 1.5 & 9.091 & 9.501 & 0.50 & 7.896 & 0.221/ & 0.197 & 0.213 
& -1.455 & 4.130 \\
& & & & & & & 0.221 & & & & \\

\end{tabular}
\end{center}
\end{table*}

\begin{table*}
\begin{center}
\begin{tabular}{clrrrrrlrrcc}
\multicolumn{12}{c}{Table 1b Cooling track characteristics} \\
\multicolumn{12}{c}{} \\
\hline
\multicolumn{12}{c}{} \\
model & {$ P_{i}$} & \multicolumn{1}{c}{$M_{i}$} & {$\lg t_{cool}$}
& \multicolumn{1}{c}{$\lg t_{evol}$} & {$X_{surf}^{f}$} & 
\multicolumn{1}{c}{$P_{f}$} & \multicolumn{1}{c}{$M_{f}$} & {$M_{2,He}(RLOF)$} 
& {$M_{2,He}(COOL)$} & {$\lg L_{f}$} &  {$\lg T_{eff,f}$} \\
& {[days]} & [{$M_\odot$}] & \multicolumn{1}{c}{[yrs]} 
& \multicolumn{1}{c}{[yrs]}& & {[days]} & [{$M_\odot$}] 
& \multicolumn{1}{c}{[$M_\odot$]} & 
\multicolumn{1}{c}{[$M_\odot$]} & [{$L_\odot$}] & \\
\multicolumn{12}{c}{} \\
\hline
\multicolumn{12}{c}{} \\
\multicolumn{12}{c}{Z=0.02} \\
\multicolumn{12}{c}{} \\

33 & 1.20 & 1.0 & 10.237 & 10.467 & 0.37 & 0.416 & 0.162 & 0.128 & 0.145 
& -2.673 & 3.830 \\
34 & 1.50 & 1.0 & 9.850 & 10.277 & 0.41 & 1.489 & 0.179 & 0.149 & 0.166 
& -2.245 & 3.929 \\
35 & 2.00~~H* & 1.0 & 9.582 & 10.193 & 0.43 & 2.912 & 0.192/ & 0.164 & 0.181
& -1.978 & 3.995 \\
& & & & & & & 0.191 & & & & \\
36 & 2.50~~H* & 1.0 & 9.420 & 10.159 & 0.45 & 4.242 & 0.200/ & 0.175 & 0.190
& -1.819 & 4.033 \\
& & & & & & & 0.199 & & & & \\
37 & 3.00~~H* & 1.0 & 9.299 & 10.139 & 0.47 & 5.551 & 0.206/ & 0.182 & 0.196
& -1.698 & 4.062 \\
& & & & & & & 0.205 & & & & \\
\multicolumn{12}{c}{} \\

38 & 1.20 & 1.5 & 10.168 & 10.255 & 0.44 & 0.736 & 0.170 & 0.141 & 0.155 
& -2.570 & 3.854 \\
39 & 1.50 & 1.5 & 9.835 & 9.987 & 0.44 & 1.737 & 0.183 & 0.154 & 0.171 
& -2.198 & 3.939 \\
40 & 2.00~~H* & 1.5 & 9.448 & 9.747 & 0.48 & 4.230 & 0.203/ & 0.179 & 0.193 
& -1.831 & 4.032 \\
& & & & & & & 0.202 & & & & \\
41 & 2.50~~H* & 1.5 & 9.295 & 9.677 & 0.49 & 5.910 & 0.210/ & 0.188 & 0.202 
& -1.690 & 4.067 \\
& & & & & & & 0.209 & & & & \\
42 & 3.00~~H* & 1.5 & 9.074 & 9.599 & 0.52 & 7.686 & 0.216/ & 0.196 & 0.209 
& -1.537 & 4.100 \\
& & & & & & & 0.215 & & & & \\
\multicolumn{12}{c}{} \\
\hline
\multicolumn{12}{c}{} \\
\multicolumn{12}{c}{Z=0.03} \\
\multicolumn{12}{c}{} \\
43 & 1.15 & 1.0 & 10.287 & 10.553 & 0.37 & 0.305 & 0.160 & 0.130 & 0.143 
& -2.753 & 3.809 \\
44 & 1.30 & 1.0 & 10.104 & 10.462 & 0.38 & 0.882 & 0.169 & 0.140 & 0.156 
& -2.562 & 3.856 \\
45 & 1.50 & 1.0 & 9.909 & 10.384 & 0.40 & 1.488 & 0.177 & 0.149 & 0.166 
& -2.343 & 3.906 \\
46 & 1.65 & 1.0 & 9.809 & 10.353 & 0.41 & 1.884 & 0.182 & 0.156 & 0.171
& -2.242 & 3.930 \\
47 & 1.80~~H* & 1.0 & 9.693 & 10.323 & 0.41 & 2.303 & 0.185/ & 0.160 & 0.175
& -2.149 & 3.950 \\
& & & & & & & 0.184 & & & & \\
48 & 2.50~~H* & 1.0 & 9.474 & 10.281 & 0.45 & 4.222 & 0.197/ & 0.175 & 0.188
& -1.912 & 4.008 \\
& & & & & & & 0.196 & & & & \\
49 & 3.00~~H* & 1.0 & 9.366 & 10.265 & 0.47 & 5.541 & 0.203/ & 0.181 & 0.195
& -1.798 & 4.035 \\
& & & & & & & 0.202 & & & & \\
\multicolumn{12}{c}{} \\
50 & 1.35 & 1.5 & 10.264 & 10.353 & 0.42 & 0.497 & 0.165 & 0.137 & 0.150 
& -2.694 & 3.822 \\
51 & 1.50 & 1.5 & 10.045 & 10.173 & 0.42 & 1.190 & 0.174 & 0.146 & 0.161 
& -2.464 & 3.876 \\
52 & 1.70 & 1.5 & 9.829 & 10.015 & 0.42 & 1.968 & 0.184 & 0.156 & 0.173 
& -2.246 & 3.929 \\
53 & 1.80~~H* & 1.5 & 9.580 & 9.873 & 0.46 & 3.380 & 0.196/ & 0.174 & 0.187
& -2.020 & 3.984 \\
& & & & & & & 0.194 & & & & \\
54 & 2.50~~H* & 1.5 & 9.353 & 9.772 & 0.49 & 5.850 & 0.208/ & 0.188 & 0.200
& -1.801 & 4.039 \\
& & & & & & & 0.207 & & & & \\
55 & 3.00~~H* & 1.5 & 9.151 & 9.705 & 0.51 & 7.671 & 0.214/ & 0.195 & 0.207
& -1.651 & 4.073 \\
& & & & & & & 0.213 & & & & \\
\end{tabular}
\end{center}
\begin{flushleft}
{Listed are:\\
$\rm P_i$ is initial orbital period of the system (at the beginning of mass transfer)\\
 $\rm M_i$ is the mass of the progenitor of white dwarf \\
$\rm t_{cool}$ is duration of the cooling phase of a white dwarf starting at the end of
RLOF\\
$\rm t_{evol} $ is total evolution time \\
 $\rm X_{surf}^{f} $ is  the final surface hydrogen content\\
$\rm P_f$ is final orbital period at the moment of shrinking of the donor within its Roche lobe\\
$\rm M_f$ is  final WD mass \\ 
$\rm M_{2,He}(RLOF)$  is the    mass of the helium core at the moment 
of shrinking of the donor within its Roche lobe\\
 $\rm M_{2,He}(COOL)$ is
the final mass of helium core after the central temperature has decreased 
by 50\% of its maximum value\\
$L_f$ is the final luminosity\\
 $T_{eff,f}$ is  the final effective temperature.\\ }
{H}{~- hydrogen flashes without RLOF}\\
{H*}{~- hydrogen flashes with RLOF}\\
\end{flushleft}
\end{table*}

\begin{table*}
\begin{center}
\begin{tabular}{lcccccc}
\multicolumn{7}{c}{Table 2 M-R relation for a cooling low--mass WD with a 
helium core} \\
\hline
\multicolumn{7}{c}{} \\
$\rm M_{wd}/M_\odot $ & $\rm R_0/R_\odot $ & $\rm R_{8500}/R_0 $ & 
$\rm R_{8500}/R_0 $ & log $\rm g_1 $ & $\rm R_{8500}/R_0 $ & log $\rm g_2 $ \\
\multicolumn{6}{c}{} \\
\hline
0.155     & 0.0218 & 2.100 & -     & 6.31 & 1.351 & 6.69 \\
0.180     & 0.0208 & 1.594 & 1.687 & 6.65 & 1.300 & 6.82 \\
0.206     & 0.0198 & 1.469 & 1.476 & 6.83 & 1.236 & 7.00 \\
0.296$^*$ & 0.0173 & 1.224 & 1.220 & 7.26 & 1.111 & 7.36 \\
\hline
\end{tabular}
\end{center}
{The first two columns present the zero--temperature
M--R relation for a helium WD obtained by Hamada \& Salpeter (1961). 
The third and fifth columns display our calculations  of the 
stellar radius and gravity, while fourth and fifth the DSBH98 calculations, 
respectively. The last two columns 
illustrate the same quantities taken from the cooling tracks 
produced by Wood (1990) for carbon WDs with thick hydrogen envelopes. 
The stellar radius is calculated at T = 8500 K and is normalized by 
the zero--temperature radius.}\\

{*}{the last two values in this row are taken from IT86.} 
\end{table*}

\begin{table*}
\begin{center}
\begin{tabular}{cccc}
\multicolumn{4}{c}{Table 3 Comparison of the cooling time-scales of} \\
\multicolumn{4}{c}{HP98, BA98 and Webbink, ours models} \\
\hline
\multicolumn{2}{c}{} & \multicolumn{1}{c}{} & \multicolumn{1}{c}{} \\
\multicolumn{2}{c}{} & \multicolumn{1}{c}{HP98 and BA98} & 
\multicolumn{1}{c}{Webbink and ours} \\
\multicolumn{2}{c}{} & \multicolumn{1}{c}{} & \multicolumn{1}{c}{} \\
\hline
& & & \\
{$\rm M_{He}/M_\odot$} & $\rm \log L/L_\odot $ & {$\rm t_{cool}$ (Gyrs)} & 
{$\rm t_{cool}$} (Gyrs) \\
& & & \\
\hline
0.15 & -3.1 & 1.0 & {36.4} \\
0.25 & -2.9 & 1.0 &  {6.1} \\
0.30 & -2.9 & 1.0 & {4.2} \\
\hline
\end{tabular}
\end{center}
\end{table*}

\begin{table*}
\begin{center}
\begin{tabular}{ccrrrrccrrr}
\multicolumn{11}{c}{Table 4a Flash characteristics} \\
\multicolumn{11}{c}{} \\
\hline
\multicolumn{11}{c}{} \\
model & case & \multicolumn{1}{c}{$\lg \Delta t_1$} 
& \multicolumn{1}{c}{$\lg \Delta t_{acc}$} 
& \multicolumn{1}{c}{$\lg \Delta t_2$} & \multicolumn{1}{c}{$\lg \Delta T$} 
& $\lg L_{max}/L_\odot$ & $\lg T_{eff}$ & $M_{b,env}$ & $M_{a,env}$
& \multicolumn{1}{c}{$\Delta M_{acc}$} \\
& & [yrs] & [yrs] & [yrs] & [yrs] & & [L=$L_{max}$] 
& \multicolumn{1}{c}{[$M_\odot$]} & \multicolumn{1}{c}{[$M_\odot$]} 
& \multicolumn{1}{c}{[$\times 10^{-4}M_\odot$]} \\ 
\multicolumn{11}{c}{} \\
\hline
\multicolumn{11}{c}{} \\
7 & 1 & 6.377 & & 6.349 & 6.595 & 0.526 & 4.349 & & & \\
& 1 & 6.852 & & 6.599 & 6.450 & 1.525 & 4.117 & & & \\
& 1 & 7.095 & & 6.609 & 6.584 & 1.603 & 4.073 & & & \\
& 1 & 7.346 & & 6.598 & & 1.693 & 4.011 & & & \\
\multicolumn{11}{c}{} \\
8 & 1 & 6.453 & & 6.509 & 6.520 & 1.571 & 4.147 & & & \\
& 1 & 6.906 & & 6.580 & 6.553 & 1.658 & 4.084 & & & \\
& 1 & 7.183 & & 6.596 & & 1.740 & 4.037 & & & \\
\multicolumn{11}{c}{} \\
12 & 1 & 6.634 & & 6.551 & 6.484 & 0.912 & 4.330 & & & \\
& 1 & 6.907 & & 6.608 & 6.449 & 1.548 & 4.084 & & & \\
& 1 & 7.139 & & 6.601 & 6.649 & 1.619 & 4.044 & & & \\
& 2 & 7.367 & 2.673 & 6.456 & & 1.703 & 4.015 & 0.0129 & 0.0123 & 1.7 \\
\multicolumn{11}{c}{} \\
13 & 1 & 6.717 & & 6.590 & 6.474 & 1.577 & 4.099 & & & \\
& 1 & 6.971 & & 6.612 & 6.574 & 1.646 & 4.053 & & & \\
& 2 & 7.215 & 2.192 & 6.571 & & 1.727 & 3.979 & 0.0119 & 0.0118 & 0.2 \\
\multicolumn{11}{c}{} \\
14 & 1 & 6.372 & & 6.519 & 6.473 & 1.588 & 4.119 & & & \\
& 1 & 6.861 & & 6.585 & 6.475 & 1.670 & 4.078 & & & \\
& 1 & 7.141 & & 6.588 & 7.603 & 1.751 & 3.994 & & & \\
& 2 & 6.435 & 2.593 & 6.258 & & 1.843 & 3.982 & 0.0113 & 0.0105 & 3.5 \\
\multicolumn{11}{c}{} \\
15 & 1 & 6.538 & & 6.558 & 6.443 & 1.669 & 4.099 & & & \\
& 1 & 6.975 & & 6.567 & 6.516 & 1.748 & 4.038 & & & \\
& 2 & 7.283 & 2.433 & 6.450 & & 1.842 & 3.963 & 0.0108 & 0.0103 & 1.2 \\
\multicolumn{11}{c}{} \\
16 & 1 & 6.643 & & 6.551 & 6.462 & 1.757 & 4.082 & & & \\
& 1 & 7.091 & & 6.553 & 6.622 & 1.846 & 4.016 & & & \\
& 2 & 7.501 & 2.513 & 6.172 & & 1.965 & 3.983 & 0.0100 & 0.0095 & 4.4 \\
\multicolumn{11}{c}{} \\
17 & 1 & 5.774 & & 6.646 & 6.510 & 1.829 & 4.047 & & & \\
& 2 & 7.202 & 2.272 & 6.477 & 6.723 & 1.933 & 3.940 & 0.0120 & 0.0094 & 0.6 \\
& 2 & 7.688 & 2.389 & 5.931 & & 2.094 & 3.983 & 0.0094 & 0.0082 & 8.2 \\
\multicolumn{11}{c}{} \\
22 & 1 & 6.495 & & 6.513 & 6.674 & 0.312 & 4.253 & & & \\
& 2 & 7.029 & 3.154 & 6.113 & & 1.311 & 3.938 & 0.0121 & 0.0116 & 4.7 \\
\multicolumn{11}{c}{} \\
23 & 2 & 6.998 & 2.994 & 6.124 & 6.884 & 1.459 & 3.922 & 0.0115 & 0.0105 
& 4.9 \\
& 2 & 7.902 & 2.806 & 5.915 & & 1.679 & 3.975 & 0.0105 & 0.0088 & 14.1 \\
\multicolumn{11}{c}{} \\
24 & 2 & 6.823 & 2.914 & 6.171 & 6.801 & 1.546 & 3.914 & 0.0103 & 0.0097 
& 3.9 \\
& 2 & 7.623 & 2.769 & 5.966 & & 1.695 & 3.939 & 0.0097 & 0.0086 & 10.4 \\
\multicolumn{11}{c}{} \\
29 & 1 & 6.662 & & 6.583 & 6.634 & 0.371 & 4.246 & & & \\
& 2 & 6.971 & 3.177 & 5.984 & & 1.307 & 3.293 & 0.0120 & 0.0108 & 6.2 \\
\multicolumn{11}{c}{} \\
30 & 2 & 6.880 & 3.066 & 5.967 & & 1.412 & 3.918 & 0.0112 
& 0.0100 & 5.7 \\
\multicolumn{11}{c}{} \\
31 & 2 & 6.652 & 2.994 & 6.011 & 6.793 & 1.470 & 3.906 & 0.0103 
& 0.0095 & 4.3 \\
& 2 & 7.666 & 2.843 & 5.899 & & 1.641 & 3.955 & 0.0095 & 0.0083 & 11.6 \\
\multicolumn{11}{c}{} \\
32 & 2 & 7.005 & 2.779 & 5.825 & & 1.682 & 3.900 & 0.0090 & 0.0078 & 7.9 \\

\end{tabular}
\end{center}
\end{table*}

\begin{table*}
\begin{center}
\begin{tabular}{ccrrrrccrrr}
\multicolumn{11}{c}{Table 4b Flash characteristics} \\
\multicolumn{11}{c}{} \\
\hline
\multicolumn{11}{c}{} \\
model & case & \multicolumn{1}{c}{$\lg \Delta t_1$} 
& \multicolumn{1}{c}{$\lg \Delta t_{acc}$} 
& \multicolumn{1}{c}{$\lg \Delta t_2$} & \multicolumn{1}{c}{$\lg \Delta T$} 
& $\lg L_{max}/L_\odot$ & $\lg T_{eff}$ & $M_{b,env}$ & $M_{a,env}$
& \multicolumn{1}{c}{$\Delta M_{acc}$} \\
& & [yrs] & [yrs] & [yrs] & [yrs] & & [L=$L_{max}$] 
& \multicolumn{1}{c}{[$M_\odot$]} & \multicolumn{1}{c}{[$M_\odot$]} 
& \multicolumn{1}{c}{[$\times 10^{-4}M_\odot$]} \\ 
\multicolumn{11}{c}{} \\
\hline
\multicolumn{11}{c}{} \\
35 & 1 & 6.510 & & 6.425 & 6.799 & -0.068 & 4.205 & & & \\
& 2 & 6.864 & 3.290 & 5.969 & & 1.173 & 3.911 & 0.0111 & 0.0102 & 7.4 \\
\multicolumn{11}{c}{} \\
36 & 2 & 6.720 & 3.158 & 5.850 & & 1.313 & 3.897 & 0.0102 & 0.0089 
& 8.6 \\
\multicolumn{11}{c}{} \\
37 & 2 & 6.885 & 2.999 & 5.810 & & 1.419 & 3.877 & 0.0097 
& 0.0082 & 10.5 \\
\multicolumn{11}{c}{} \\
40 & 2 & 6.622 & 3.168 & 5.549 & & 1.275 & 3.883 & 0.0098 & 0.0084 & 9.2 \\
\multicolumn{11}{c}{} \\
41 & 2 & 6.876 & 2.981 & 5.719 & & 1.427 & 3.875 & 0.0087 & 0.0076 & 10.2 \\
\multicolumn{11}{c}{} \\
42 & 2 & 7.362 & 2.744 & 5.737 & & 1.615 & 3.876 & 0.0079 & 0.0063 & 11.6 \\
\multicolumn{11}{c}{} \\
47 & 1 & 7.373 & & 6.594 & 6.966 & 0.208 & 4.151 & & & \\
& 2 & 6.095 & 3.388 & 5.938 & & 1.072 & 3.926 & 0.0109 & 0.0095 & 12.7 \\
\multicolumn{11}{c}{} \\
48 & 2 & 6.809 & 3.200 & 5.820 & & 1.239 & 3.864 & 0.0094 
& 0.0080 & 11.1 \\
\multicolumn{11}{c}{} \\
49 & 2 & 6.871 & 3.081 & 5.836 & & 1.317 & 3.863 & 0.0089 & 0.0074 & 9.9 \\
\multicolumn{11}{c}{} \\
53 & 1 & 6.471 & & 6.690 & 6.819 & 0.190 & 4.170 & & & \\
& 2 & 6.920 & 3.239 & 6.785 & & 1.160 & 3.889 & 0.0094 & 0.0078 & 12.4 \\
\multicolumn{11}{c}{} \\
54 & 2 & 6.927 & 2.962 & 5.722 & & 1.355 & 3.855 & 0.0081 & 0.0066 & 11.1 \\
\multicolumn{11}{c}{} \\
55 & 2 & 7.180 & 2.785 & 5.734 & & 1.521 & 3.849 & 0.0074 & 0.0058 & 11.8 \\

\end{tabular}
\end{center}
\begin{flushleft}
{Listed are:\\
 number of model (Table 1)\\
 number of case (1 or 2)\\  
$\Delta t_1$ and $\Delta t_2$ are the rise and decay times responsively\\ 
$\Delta T$ and $\Delta t_{acc}$ are recurrence time between two successful 
flashes and duration of accretion phase during the flash \\
  $T_{eff}$ is effective temperature when the luminosity has its maximum value $L_{max} $\\ 
$M_{b,env}$ and 
$M_{a,env}$ are the envelope masses before and after flash \\
 $\Delta M_{acc}$ is accreted mass\\ 
 $M_{b,env}$-$M_{a,env}$=
$\Delta M_{He,c}$+$\Delta M_{acc}$ where $\Delta M_{He,c}$ is the increase 
of the helium core mass during the flash.\\}
\end{flushleft}
\end{table*}

\end{document}